\documentstyle[prl,aps,twocolumn]{revtex}

\begin{document}

\title{Inflationary cosmology - a dissipative quantum field theory process}

\author{Arjun Berera\thanks{PPPAC Advanced Fellow}}

\address{
{\it School of Physics,
University of Edinburgh, Edinburgh EH9 3JZ, Great Britain} 
\\
{\rm To appear in Proceedings ICHEP02, Amsterdam, July 2002}
}

\maketitle

\begin{abstract}
The current status of early universe cosmology can be summarized as
precision, but inherently limited, data requiring 
explanation from precision and
robust theory.  Specifically inflation is presently the single
concrete hope for solving the early universe problem, since it
is generically realizable in quantum field theory, but
treatment of interactions needs considerable attention.
This talk reviews the importance and progress in this direction.

\vspace{1pc}
\end{abstract}

\bigskip

\noindent

\bigskip

% typeset front matter (including abstract)
\maketitle

\section{Introduction}

Early universe cosmology is a subject with limited access
to observation, since there is only so much information that can be
acquired about this time period from our one local region
in the universe.  As such, a convincing theory about this
time period must rely substantially on the soundness and
unambiguity of its mathematical foundation.
At present, inflation is the most hopeful idea for developing
into a theory, since it is based on quantum field theory,
and quantum fields so far have been 
exclusively successful in explaining the
high energy world, up to at least the TeV scale. 
However, consistent solutions of inflation founded
on quantum field theory have been allusive. 
Inflationary dynamics inherently is a multifield problem,
since the vacuum energy that drives inflation eventually
must convert to radiation, which generally
is comprised of a variety of particle species. 
In the earliest conception of inflation \cite{si}, it was
pictured that inflation would result in a isentropic
expansion that would rapidly put the universe in
a supercooled thermodynamic phase.  
Subsequently it was observed that supercooled
inflation is not mandatory and that nonisentropic 
inflationary expansion, warm inflation, also is
possible \cite{wi}.
Moreover, since the main condition for inflation is
$\rho_v > \rho_r$, where $\rho_v$,$\rho_r$ are the vacuum
and radiation energy densities respectively, supercooled inflation
appears as a limiting
case within the general regime of warm inflation.

The most nontrivial aspect of the inflaton model is understanding the
energy transfer dynamics from potential energy to radiation.
In the most commonly followed inflation picture, dissipative effects
of the inflaton field are ignored throughout
the inflation period, leading to a supercooled inflationary
regime.  However, from a thermodynamic perspective, this
picture appears very restrictive. The point being, even if
the inflaton were to allow a minuscule fraction
of the energy to be released, say one part in $10^{20}$,
it still would constitute a significant radiation
energy density component in the universe.
{}For example, for inflation with vacuum (i.e. potential) energy
at the GUT scale $\sim 10^{15-16} {\rm GeV}$,
leaking one part in $10^{20}$ of this energy density into radiation 
corresponds to a temperature of $10^{11} {\rm GeV}$,
which is nonnegligible.  In fact, the most relevant
lower bound that cosmology places on the temperature after inflation
comes from the success of hot Big-Bang nucleosynthesis,
which requires the universe to be within the
radiation dominated regime by $T \stackrel{>}{\sim} 1 {\rm GeV}$.
This limit can be met in the above example by dissipating 
as little as one part in $10^{60}$ of the vacuum energy
density into radiation.
Thus, from the perspective of both interacting field theory and
basic notions of equipartition,  it appears to be a highly
tuned requirement of supercooled inflation to prohibit
the inflaton from such tiny amounts of dissipation.

These considerations have led to examining the possibility
of warm inflation, an inflationary regime in which radiation
also is present.
Warm inflation is comprised of non-isentropic expansion in the
background cosmology \cite{wi} and thermal seeds of density
perturbations \cite{bf2,wipert} (for related earlier work
please see \cite{moss}).  
During warm inflation,
interactions between the inflaton
and other fields cause
the radiation energy
density to remain substantial due to its constant production from
conversion of vacuum energy.  This expansion regime is intrinsically
different from the supercooled inflation regime, since warm
inflation smoothly terminates into a
subsequent radiation dominated regime, without a reheating period.

The warm inflation picture has one immediate conceptual
advantage, that the dynamics is
free of questions about quantum-to-classical
transition. The scalar inflaton field is in a
classical state, thus justifying the application
of  a classical evolution equation, and the inflaton fluctuations,
which induce the metric perturbations, are
classical.  Another problem alleviated by dissipative effects
is the initial condition problem 
of inflation \cite{bg}.

Warm inflation continues proving to be a promising solution to
the early universe problem, with now several
variants and perspectives on the basic idea \cite{wipert,wimodels}. 
Development of it has had two directions,
which will be addressed in this talk.
First is deriving a first principle quantum field theory realization
of its dynamics. Second is determining observational tests
for warm inflation.

\section{First Principles Origin}

The earliest works looked for high temperature
warm inflation solutions, under rigid adiabatic, equilibrium
conditions \cite{bgr}.  Within this limited framework, 
one type of warm inflation solution
was obtained \cite{bgr}. 
The high-T regime was examined first, since considerable
methodology was already available for treating it. 
However, intrinsically, the statistical state relevant for
warm inflation is not required to be an equilibrium
state.   The slowly varying nature of  the macroscopic variables
in warm inflation cosmology suggest that the statistical state
may not be far from equilibrium, although this is something
that should be proven from the dynamics.  Much work remains
in order to develop the mathematical formalism necessary to
address this problem.
As one step in this direction to fill the missing gaps, recently we
studied the zero temperature dissipative
dynamics of interacting scalar field 
systems in Minkowski spacetime \cite{br} (for another interesting
direction see \cite{idl}).
This is useful to understand, since
the zero temperature limit constitutes a baseline effect that
will be prevalent in any general statistical state.
The key result presented in this talk is that
for a broad range of cases, involving interaction
with as few as one or two fields, we find dissipative regimes
for the scalar field system.  This is important for inflationary
cosmology, since it suggests that dissipation
may be the norm not exception for an interacting scalar field system,
thus warm
inflation could be a natural dynamics once proper 
treatment of
interactions is done.

We study the Lagrangian, where
\begin{eqnarray} 
{\cal L}_I &=&  
-\frac{\lambda}{4 !} \Phi^4  
- \sum_{j=1}^{N_{\chi}} \left\{ 
\frac{f_{j}}{4!} \chi_{j}^4 + \frac{g_{j}^2}{2} 
\Phi^2 \chi_{j}^2  
\right\} \nonumber \\ 
& - & \sum_{k=1}^{N_{\psi}}   
\bar{\psi}_{k} \left[h_{k\phi} \Phi 
+ \sum_{j=1}^{N_\chi} h_{kj\chi} \chi_j \right] \psi_{k} 
\:
\label{Nfields} 
\end{eqnarray} 
and $\Phi \equiv \varphi + \phi$ such that
$\langle \Phi \rangle = \varphi$.  Our aim is to
obtain the effective equation of motion for $\varphi(t)$ and
from that determine the energy dissipated from the $\varphi(t)$
system into radiation.

The tadpole method, which requires 
$\langle \phi \rangle =0$,
gives effective equation of motion for $\varphi(t)$,
\begin{eqnarray} 
\ddot{\varphi}(t) + m_\phi^2 \varphi(t) + \frac{\lambda}{6} \varphi^3(t)  
+\frac{\lambda}{2} \varphi(t) \langle \phi^2 \rangle
+\frac{\lambda}{6} \langle \phi^3 \rangle  \nonumber \\
+ \sum_{j=1}^{N_{\chi}} g_j^2 \left[\varphi (t) \langle \chi_j^2 \rangle + 
\langle \phi \chi_j^2 \rangle \right]
+ \sum_{k=1}^{N_{\psi}} h_{k\phi} \langle \bar{\psi_k} \psi_k \rangle= 0 \;. 
\label{eq2phi1} 
\end{eqnarray} 
The field expectation values in this equation are
obtained by solving the coupled set of field equations.
In our calculation, we have evaluated them in a perturbative
expansion using dressed Green's functions. 
One general feature of these expectation values
is they will depend of the causal history of $\varphi(t)$,
so that Eq. (\ref{eq2phi1}) is a temporally nonlocal equation
of motion for $\varphi(t)$.  

The general expression for the effective equation
of motion is given in \cite{br} and is very complicated.
{}Formally, we can examine Eq. (\ref{eq2phi1})
within a Markovian-adiabatic approximation, in which
the equation of motion is local in time and the motion
of $\varphi(t)$ is slow.  At $T=0$, such an approximation
is not rigorously valid.  Nevertheless, this approximation
allows understanding the magnitude of dissipative effects.
{}Furthermore, we have shown in \cite{br} that the nonlocal
effects tend to filter only increasingly higher
frequency components of $\varphi(t)$
from nonlocal effects increasingly further back in time.  
Thus for low
frequency components of $\varphi(t)$, memory only is
retained to some short interval in the past.   Since within
the adiabatic approximation, $\varphi(t)$ only has low frequency 
components, we believe the Markovian-adiabatic
approximation is legitimate at least for order of
magnitude estimates. Within this approximation,
the effective equation of motion for $\varphi(t)$ has the general
form
%\begin{eqnarray} 
\begin{equation} 
\ddot{\varphi}(t) + m_{\phi}^2 \; \varphi (t) +  
\frac{\lambda}{6} \varphi^3 (t) + \eta (\varphi)  
\dot{\varphi} (t) =0\;, 
\label{final1} 
\end{equation} 
%\end{eqnarray} 
where
%\begin{equation*} 
\begin{eqnarray*} 
\eta(\varphi)  \approx 
\varphi^2 (t) \frac{N_{\psi}}{256\pi^2}\left(
\frac{\lambda^2 h_{\phi\psi}^2}{4m_{\phi}} .
+ \frac{N_{\chi}g^4 h_{\chi\psi}^2}{m_{\chi}} \right) .
\label{eta2} 
\end{eqnarray*} 
%\end{equation*} 
 
An alternative to the above Lagrangian based derivation,
is the canonical derivation following formalism
developed in the mid 80's \cite{ms1}.  Although
the canonical and Lagrangian approaches should 
agree, the former is far less developed in dissipative
quantum field theory, in particular
for treating interactions.  Nevertheless, the canonical
approach provides useful insight, especially for understanding
the origin of particle creation.  {}For example, consider
in the canonical approach the expectation value 
of $\langle \chi_i^2 \rangle$,
\begin{eqnarray*}
\int \frac{d^3 q}{(2 \pi)^3 2 {\rm Re} \omega_{\bf q,\chi_j}}
\left[2 x_{\bf q,\chi_j}(t) + 2 {\rm Re} y_{\bf q,\chi_j}(t)+1\right] ,
\label{averchi}
\end{eqnarray*}
where
$x_{{\bf q},\chi_j}(t)= \langle a_{{\bf q},\chi_j}^{\dagger}(t) 
a_{{\bf q},\chi_j}(t) \rangle$
is the particle
number density and
$y_{\bf q,\chi_j}(t)= \langle a_{{\bf q},\chi_j}(t) a_{-{\bf q},\chi_j}(t) 
\rangle$ is the
off-diagonal correlation.  The evolution equations for
$x_{\bf q,\chi_j}(t)$ and $y_{\bf q,\chi_j}(t)$ can be obtained from the
field equation for $\chi_j$ to give
\begin{eqnarray}
{\dot x}_{{\bf q},\chi_j} & = & \frac{\dot{\omega}_{\bf q,\chi_j}}
{\omega_{\bf q,\chi_j}}
{\rm Re} \, y_{\bf q,\chi_j} \;, 
%\\
%\end{eqnarray*}
\nonumber \\
%\begin{equation}
\dot{y}_{\bf q,\chi_j} & = & \frac{\dot{\omega}_{\bf q,\chi_j}}
{\omega_{\bf q,\chi_j}}
\left[ x_{\bf q,\chi_j} +\frac{1}{2} \right] - 2 i 
\omega_{\bf q,\chi_j}
y_{\bf q,\chi_j} \;.
\label{diff}
%\end{equation}
\end{eqnarray}
To yield dissipation, it is noted in \cite{ms1} that 
the correlation
amongst produced particles needs to be destroyed
sufficiently rapidly and ${\dot x}_{\bf q,\chi_j}(t)$ and
$\langle \chi_j(t) \rangle$ should become local functions of time
involving $\varphi(t)$ and ${\dot \varphi}(t)$.
Based on this requirement,
\cite{ms1} asserts that $\omega_{\bf q,\chi_j}$ in
the above equation for ${\dot y}_{\bf q,\chi_j}(t)$
must have an imaginary part, which in fact should be
the $\chi_j$-particle decay width,
${\rm Im} \omega_{{\bf q},\chi_j} \propto \Gamma_{\bf q,\chi_j}$.
Applying these assumptions, the contribution to
$\eta(\varphi)$ in Eq. (\ref{final1}) from the $\chi_j$ field, say
$\eta_{\chi_j}(\varphi)$, once again is obtained
(up to O(1) factors). 
Since this approach has ad-hoc assumptions, it still
is incomplete and requires development. 
Nevertheless, the approach is interesting and
for now accepting the assumptions,
the origin of particle creation and energy conservation
are clearly seen. In particular, the particle
production rate is given by
$\int (d^3k/(2\pi)^3){\dot x}_{\bf q,\chi_j}(t)\omega_{\bf q,\chi_j}$
and similar to \cite{ms1}, it can be shown
this is equal to the vacuum energy loss rate
from the $\chi_j$ field contribution,
$\eta_{\chi_j}(\varphi) {\dot \varphi}^2$.

Returning to Eq. (\ref{final1}), estimated magnitudes
of energy production will be obtained 
in the overdamped regime of warm inflation,
%\begin{equation} 
$m^2(\phi) = m_{\phi}^2 + \lambda \varphi^2/2 < \eta^2$.
%\end{equation} 
In this regime,
the energy dissipated by the scalar field goes into  
radiation energy density $\rho_r$ at the rate
${\dot \rho}_r = -{dE_{\phi}}/{dt} = 
\eta(\varphi) {\dot \varphi}^2$.

In \cite{br} we have determined radiation production for
two cases 
%\begin{eqnarray}
${\rm (a). } \hspace{0.1cm} m(\varphi) > m_{\chi} > 2m_{\psi}$,
%\nonumber\\
${\rm (b). } \hspace{0.1cm} m_{\chi} > 2 m_{\psi} > m(\varphi)$.  
%\end{eqnarray}
To focus on
a case typical for inflation, suppose the potential
energy is at the GUT scale $V(\varphi)^{1/4} \sim 10^{15} {\rm GeV}$
and we consider the other parameters in a regime consistent
with the e-fold and density fluctuation
requirements of inflation.  Note, although this is a flat nonexpanding
spacetime analysis, since the dissipative effects will
be at subhorizon scale, one expects these estimates to give a reasonable
idea of what to expect from a 
similar calculation done in expanding spacetime.
Expressing the radiation in terms of a temperature scale as
$T \sim \rho_r^{1/4}$, we find for case (a)
$1 {\rm GeV} < T < 10^7 {\rm GeV} < H$ and for case (b)
$T \stackrel{>}{\sim} 10^{14} {\rm GeV} > H$,
where $H = \sqrt{8 \pi V/(3m_p^2)}$.

\section{Observational Tests}

Supercooled inflation has three parameters,  
related to the potential energy
magnitude $V_0$, slope $\epsilon$, and curvature $\eta$, whereas
there are four observable constraints ($\delta_H$, 
$A_g$, $n_s$, $n_g$).
This implies a redundancy in the observations and
allows for a consistency relation \cite{liddlelyth}. This is usually 
expressed as a relationship
between the tensor-to-scalar ratio and the slope of the tensor spectrum.
Warm inflation has an extra parameter, the dissipation factor,
which implies four constraints for four parameters. Hence we do not expect the
consistency relation of standard inflation 
to hold in warm inflation \cite{tb}.
Thus, to discriminate between warm and standard 
inflation, it requires a measure of
all four observables. The upcoming MAP and Planck satellite missions
should provide strong constraints on the scalar spectrum
and being equipped with polarization detectors, it is hoped
the tensor spectrum also will be measured.
Recently nongaussian effects from warm inflation models
were computed and found to be too small to measure \cite{gbhm}.

\section{Conclusion}

The key point emphasized in this talk has been that 
any typical inflaton field theory model has interactions and
so dissipation can be an unexceptional consequence.
These dissipative effects have always been ignored,
unjustifiably, in supercooled inflation, but as
shown here their effects can be nontrivial.  The most significant
effect of dissipation is to completely alter the inflationary
regime from supercooled to warm.  However, irrespective of
the inflation regime, as discussed above and in \cite{bg},
dissipative effects before the inflation
period may be important in alleviating the initial condition
problem.  Also,
dissipative effects imply new realizations
of thermal inflation \cite{bk}. 
Moreover, dissipative effects can alter density
perturbations.  In this talk we discussed the case
where dissipation leads to an ideal thermalized state,
thus with thermal density perturbations. However the opposite
limit can be considered from ideal thermalization 
to that of negligible influence from
the radiation field on the density perturbations, so
that only the quantum fluctuations contribute.  Then
simple calculations show
warm inflation models would {\it not}
require fine tuning, so that scalar potentials of
reasonably large curvature would be acceptable.
The nature of density perturbations during warm inflation
still requires much investigation
in particular to understand the
interplay between radiation and the scalar field.
In conclusion, this talk discussed how dissipation during inflation
could be robust and possibly
solve inflation's most challenging dilemma, the scalar field fine tuning
problem.

\end{document}